\documentclass[twocolumn,superscriptaddress,showpacs,aps]{revtex4}
\usepackage{amsmath,amssymb,bm}
\usepackage{graphicx,epstopdf} 
\usepackage{color}
\usepackage{hyperref}

\renewcommand{\d}{{\rm d}}
\newcommand{\iChEM}{{\it i}{\rm ChEM}}

\newcommand{\up}{\uparrow}
\newcommand{\down}{\downarrow}

\newcommand{\greater}{\mbox{\tiny $>$}}
\newcommand{\lesser}{\mbox{\tiny $<$}}

\newcommand{\Opm}{\hat O_{\mbox{\tiny $\pm$}}}

\newcommand{\tS}{\mbox{\tiny S}}
\newcommand{\B}{\mbox{\tiny B}}
\newcommand{\SB}{\mbox{\tiny SB}}
\newcommand{\T}{\mbox{\tiny T}}
\newcommand{\w}{\omega}
\newcommand{\ti}{\Tilde}
\newcommand{\wti}{\widetilde}
\newcommand{\dg}{\dagger}
\newcommand{\la}{\langle}
\newcommand{\ra}{\rangle}
\newcommand{\La}{\big\la}
\newcommand{\Ra}{\big\ra}

\newcommand{\nl}{\nonumber \\}
\newcommand{\Sec}[1]{Sec.\;\ref{#1}}

\newcommand{\be}{\begin{equation}}
\newcommand{\ee}{\end{equation}}
\newcommand{\bsube}{\begin{subequations}}
\newcommand{\esube}{\end{subequations}}
\newcommand{\Eq}[1]{Eq.\,(\ref{#1})}
\newcommand{\Eqs}[1]{Eqs.\,(\ref{#1})}
\newcommand{\Fig}[1]{Fig.\,\ref{#1}}

\begin{document}

\title{Mechanism of current noise spectrum
in a nonequilibirum Kondo dot system}

\author{Hong Mao}
\affiliation{Department of Physics, Hangzhou Normal University,
 Hangzhou, Zhejiang 311121, China}

\author{Jinshuang Jin} \email{jsjin@hznu.edu.cn}
\affiliation{Department of Physics, Hangzhou Normal University,
 Hangzhou, Zhejiang 311121, China}

\author{Shikuan Wang}
\affiliation{Department of Physics, Hangzhou Dianzi University, Hangzhou 310018, China}

\author{YiJing Yan} \email{yanyj@ustc.edu.cn}
\affiliation{Hefei National Laboratory for Physical Sciences at the Microscale
\& \iChEM,
University of Science and Technology of China, Hefei, Anhui 230026, China}

\date{\today}

\begin{abstract}

 We systematically study
 the nonequilibirum Kondo mechanisms of quantum noise spectrum
 based on the accurate
dissipaton--equation--of--motion evaluations.
By comparing the noise spectra
between the equilibrium and nonequilibrium cases
and between the non-Kondo and Kondo regimes,
we identify the nonequilibrium Kondo features in
the current noise spectrum,
appearing in the region of $\w\in [-eV, eV]$.
The Kondo characteristic at
$\omega=\pm eV=\pm (\mu_{\rm L}-\mu_{\rm R})$
 display asymmetrical upturns
 and remarkable peaks in $S(\omega)$
 and $\d S(\omega)/\d\omega$, respectively.
These features are originated
from the Rabi interference of the transport current dynamics,
with the Kondo oscillation frequency of $|eV|$.
The minor but
 very distinguishable inflections,
 crossing over $\w=-eV$ to $\w=+eV$,
would be related to a sort of Kondo-Fano interference
between two Kondo resonances channels.

\end{abstract}
\maketitle

\section{Introduction}
\label{thintro}

 Shot noise of nonequilibrium quantized charge
current fluctuations
carries much rich information beyond the average current
\cite{Bla001,Imr02,Bee0337,Naz03}. 
The study of current noise in transport through mesoscopic
devices becomes a field of intensive
theoretical and experimental research.
Noise spectrum is the Fourier transformation of
two--time current--current correlation function.
The zero-frequency noise describes the
steady--state fluctuations of
the effective carrier charge that can be
either fraction \cite{Pic97162,Rez99238,Bid09236802} or integer
\cite{Koz003398,Lef03067002}.
Noise spectrum in full frequency domain contains
both static and dynamic information.
It is a powerful probe to the energetics, interactions
and dynamics of strongly correlated systems \cite{Ent07193308,Li05066803,Bar06017405,%
Gab08026601,Wab09016802,Liu131866,
Jin13025044,Jin11053704,Eng04136602,Rot09075307,Yan14115411}.

Thanks to the recent advancement in on-chip detection technique,
high-precision measurement of nonequilibrium
current fluctuations in a
Kondo quantum dot (QD) is now available at
  finite frequency \cite{Bas10166801,Bas12046802,Del18041412}.
In the finite-frequency noise spectrum, the Kondo feature is
predicted a logarithmic singularity at $\omega =\pm eV$
 \cite{Moc11201303,Mul13245115}.
One can also observe the Kondo peaks
in the derivative noise
to bias voltage $V$
 \cite{Bas12046802,Del18041412}.
 In particular, the emissive spectrum,
 which is largely uncontaminated,
has been studied experimentally \cite{Del18041412} and
theoretically \cite{Cre18107702},
for QDs in asymmetrical coupling to reservoirs.

 In this work, we will explore the nonequilibirum Kondo mechanism
 in the noise spectrum of the current tunneling through
 an Anderson impurity quantum dot.
It is well-known that the equilibrium Kondo effect leads to
the impurity density of state (DOS), $A(\omega)$,
a resonance peak, which splits into two peaks
under an applied bias voltage.
The observed Kondo resonance at Fermi level is due to the formation
of singlet on the QD screened by itinerant electrons
from reservoirs.
Away from the Kondo regime, the DOS also contains the two Hubbard resonances peaks at
the single-occupation and double occupation transport resonances.
That is, the DOS $A(\omega)$ reflects the structure information of the impurity system.
%

   On the other hand, the nonequilibrium noise spectrum,
$S(\w)$, the Fourier transformation
of current--current correlation function,
involves not only the structure
but also the transport dynamics.
We will systematically investigate
the nonequilibrium Kondo characteristics
in both $S(\omega)$ and $\d S(\omega)/\d\omega$.
 The underlying mechanisms are identified,
against the possible competing processes.
Some details are as follows.
(\emph{i})
We illustrate the Kondo characteristic in
both $S(\omega)$ and $\d S(\omega)/\d\omega$,
with a close comparison between
the equilibrium and nonequilibrium cases.
The observed Kondo characteristic is related only with
the Fermi energies difference between two electrodes.
Namely, the applied bias voltage
splits the Kondo characteristics,
from the single inflection point in $S(\omega)$
and the peak $\d S(\omega)\d\w$ at $\omega=0$,
into two asymmetric upturns and peaks, respectively,
at around $\omega=\pm|\mu_{\rm L}-\mu_{\rm R}|=\pm eV$;
(\emph{ii}) By comparing the non-Kondo and Kondo regimes,
we identify the nonequilibrium Kondo features in
the current noise spectrum,
appearing in the region of $\w\in [-eV, eV]$.
This is a type of Kondo-Fano interference,
engaging both Kondo characteristics
at $-eV$ and $+eV$.
In contrast, the non-Kondo cotunneling
process is of the anti-Stokes in nature,
occurring at $-eV$ only, which rules
out the interference in $\w\in [-eV, eV]$;
(\emph{iii}) We establish a bridge
between the nonequilibrium Kondo noise
and the transient current.
The observed Kondo profile around $\omega=\pm eV$
reflects the Rabi interference of the transport current dynamics
and contains  the information of the
Kondo oscillation frequency $|eV|$;
(\emph{iv})
We illustrate that the emission noise Kondo feature
($\w<0$) is often
cleaner than the absorption ($\w>0$), as the
latter would be contaminated by the sequential tunneling signals.

The present study is based on the well-established
dissipaton equation of motion (DEOM)
approach \cite{Jin15234108,Yan14054105,Yan16110306,Zha18780,Wan20041102}.
This is a nonperturbative and accurate method,
having been extensively explored in the study of quantum impurity problems
\cite{Jin15234108,Zha16237,Zha16204109,%
Che20297811,Wan20164113,Gon20154111,Jin20235144}.
These include the recent noise spectrum evaluations,
with the identification of Coulomb blockade assisted
Rabi interference in a double-dot
Aharonov-Bohm interferometer \cite{Jin20235144}.

The remainder of this paper is organized as follows.
In \Sec{thmet}, we give a brief introduction of
the DEOM theory and demonstrate how it evaluates the two-time
current-current correlation function.
In \Sec{thnum}, we demonstrate and elaborate the numerical
 results of the circuit current noise spectrum
and the related transient circuit current.
Finally, we conclude this work with \Sec{thsum}.

\section{Methodology}
\label{thmet}

 In this section, we present a brief account on the DEOM theory and
 current-current correlation function. For details see
 References \cite{Yan14054105,Jin15234108,Yan16110306}.
 %

 Consider an electron transport setup
in which an impurity system $H_{\tS}$ is sandwiched by electrode bath $h_{\B}$,
under an electric bias potential ($eV=\mu_{{\rm L}}-\mu_{{\rm R}}$)
applied across the leads, $\alpha = {\rm L}$ and R.
The total Hamiltonian reads $H_{\rm tot}=H_{\tS}+H_{\B}+H_{\SB}$.
The system Hamiltonian $H_{\tS}$ is arbitrary,
including electron-electron interaction,
given in terms of local electron creation $\hat a^{\dg}_{u}$
(annihilation $\hat a_{u}$) operators.
For instance, in the present study
we consider a QD represented by the spin-$1/2$
single Anderson impurity model (SAIM) described by
\be
 H_{\tS}
=\sum_{u=\up,\down} \varepsilon_{u} \hat a^\dg_u \hat a_u
+\frac{U}{2}\sum_u\hat n_{u}\hat n_{\bar u},
\ee
where the single level in the QD is characterized by a
spin-degenerate energy level $\varepsilon_{\up}=
\varepsilon_{\down}=\varepsilon$ and
 $\hat n_u=\hat  a^\dg_{u} \hat a_{u}$,
 with $\bar u$ the opposite direction of the spin index $u$.

 The electrode bath is modeled as noninteracting electrons
reservoirs,
$ H_{\B}= \sum_{\alpha k}(\varepsilon_{\alpha k}
  +\mu_{\alpha}) c^{\dg}_{\alpha k}  c_{\alpha k}$.
 Its coupling to the system
assumes the standard tunneling form of
\be\label{Hsb1}
  H_{\SB}\!=\!\sum_{\alpha u }\left(\hat a^{+}_{u}  \hat F^-_{\alpha u}
    + \hat F^+_{\alpha u} \hat a^{-}_{u} \right)\!=\!\sum_{\alpha u \sigma}\hat a^{\bar\sigma}_{u}
    \wti F^{\sigma}_{\alpha u},
\ee
with $ \hat F^-_{\alpha u}
=\sum_k t_{\alpha u k} c_{\alpha k}=(\hat F^+_{\alpha u})^\dg$.
Note that $\hat F^{\sigma}_{\alpha u}\hat a^{\bar\sigma}_{u}
=-\hat a^{\bar\sigma}_{u}\hat F^{\sigma}_{\alpha u}$.
For the convenience of description, we denote
 $\wti F^{\sigma}_{\alpha u} \equiv \bar\sigma \hat  F^{\sigma}_{\alpha u}$,
with $\sigma =+,-$ ($\bar\sigma$ is the opposite sign) 
identifying the creation and annihilation operators.

For the dissipaton description of bath interaction \cite{Jin15234108,Yan14054105},
we consider the bare-bath correlation function
in the exponential decomposition form \cite{Jin08234703,Zhe121129,
Zhe13086601,Li12266403,Hou15104112,Ye16608},
\be\label{FF_corr}
 \big\la \hat F^{\sigma}_{\alpha u}(t)
\hat F^{\bar\sigma}_{\alpha v}(0)\big\ra_{\B}
 = \sum_{m=1}^{M} \eta^{\sigma}_{\alpha uv m}e^{-\gamma^{\sigma}_{\alpha m}t}.
\ee
This is realized via a sum-over-poles decomposition
for the Fourier integrand of the relation
  $\la \hat F^{\sigma}_{\alpha u}(t)\hat F^{\bar\sigma}_{\alpha v}(0)\ra_{\B}
=\frac{1}{\pi}\int_{-\infty}^{\infty}\!\!d\omega\,e^{\sigma i(\omega+\mu_{\alpha}\!)t}
   \frac{J^{\sigma}_{\alpha uv}(\omega)}
    {1+e^{\sigma\beta\omega}}$.
Here, $J^+_{\alpha vu}(\w) = J^-_{\alpha uv}(\w) = J_{\alpha uv}(\w)$;
 the reservoir hybridization spectral function is given by
 $J_{\alpha u v}(\omega)
\equiv\pi\sum_k t_{\alpha u k}t^\ast_{\alpha v k}\delta(\omega-\varepsilon_{\alpha k})
=\frac{\Gamma_{\alpha u v}W^2}{\omega^2+W^2}$.
The exponents $\{\gamma^{\sigma}_{\alpha m}\}$ in \Eq{FF_corr}
arise from both the Fermi function
and the hybridization function.
For an optimal dissipaton description, we adopt the Pad\'{e} spectrum decomposition
for Fermi function \cite{Hu10101106,Hu11244106}.

 The DEOM theory starts with the
statistical quasi--particle (dissipaton) decomposition
on the hybridizing bath operators $\{\hat F^{\sigma}_{\alpha u}\}$.
It reproduces the bath correlation functions,
\Eq{FF_corr}, and the time--reversal
counterparts,
$\big\la \hat F^{\bar\sigma}_{\alpha v}(0)
 \hat F^{\sigma}_{\alpha u}(t)\big\ra_{\B}
=\La\hat F^{\bar\sigma}_{\alpha u}(t)
 \hat F^{\sigma}_{\alpha v}(0)\Ra^{\ast}_{\B}.
$
To that end, we set \cite{Yan14054105,Jin15234108,Yan16110306}
\be\label{wtiF_f}
  \wti F^{\sigma}_{\alpha u} \equiv -\sigma \hat  F^{\sigma}_{\alpha u}
    \equiv \sum_{m=1}^{M} \hat f^{\sigma}_{\alpha u m} .
\ee
The involved dissipatons $\{f^{\sigma}_{\alpha u m}\}$ satisfy
\be\label{ff_corr}
\begin{split}
  \big\la\hat f^{\sigma}_{\alpha u m}(t)\hat f^{\sigma'}_{\alpha' v m'}(0)\big\ra_{\B}
 =\big\la\hat f^{\sigma}_{\alpha u m}\hat f^{\sigma'}_{\alpha' v m'}\big\ra^{\greater}_{\B}\, e^{-\gamma^{\sigma}_{\alpha m} t},
\\
 \big\la\hat f^{\sigma'}_{\alpha' v m'}(0)\hat f^{\sigma}_{\alpha u m}(t)\big\ra_{\B}
 =\big\la\hat f^{\sigma'}_{\alpha' v m'}\hat f^{\sigma}_{\alpha u m}\big\ra^{\lesser}_{\B}\, e^{-\gamma^{\sigma}_{\alpha m} t},
\end{split}
\ee
where $\gamma^{\bar\sigma\,\ast}_{\alpha m}=\gamma^{\sigma}_{\alpha m}$ and
 \be\label{ff_corr0}
\begin{split}
   \big\la\hat f^{\sigma}_{\alpha u m}\hat f^{\sigma'}_{\alpha' v m'}\big\ra^{\greater}_{\B}
   =-\delta_{\sigma\bar\sigma'}\delta_{\alpha\alpha'}\delta_{mm'}\,
   \eta^{\sigma}_{\alpha u v m},
\\
  \big\la\hat f^{\sigma'}_{\alpha' v m'}\hat f^{\sigma}_{\alpha u m}\big\ra^{\lesser}_{\B}
  =-\delta_{\sigma\bar\sigma'}\delta_{\alpha\alpha'}\delta_{mm'}\,
   \eta^{\bar\sigma\,\ast}_{\alpha u v m}.
\end{split}
\ee

 For bookkeeping, we adopt the abbreviations,
 $j\equiv(\sigma\alpha u m)$ and $\bar j\equiv(\bar\sigma\alpha u m)$,
for the collective indexes in fermionic dissipatons, such
that $f_j\equiv f^{\sigma}_{\alpha u m}$ and so on.
Dynamical variables in DEOM are the reduced dissipaton density
operators (DDOs),
\be\label{DDO_def}
 \rho^{(n)}_{\bf j}(t)\equiv \rho^{(n)}_{j_1\cdots j_n}(t)\equiv
 {\rm tr}_{\B}\Big[\big(\hat f_{j_n}\cdots\hat f_{j_1}\big)^{\circ}
  \rho_{\rm tot}(t)\Big]\, .
\ee
The product of dissipatons inside the circled parentheses, $(\,\cdot\cdot\,)^{\circ}$,
is \emph{irreducible}.
A swap of any two irreducible fermionic dissipatons
causes a minus sign, such that
 $\big(\hat f_{j}\hat f_{j'}\big)^{\circ}=-\big(\hat f_{j'}\hat f_{j}\big)^{\circ}$.
While $\rho_{\tS}(t) \equiv \rho^{(0)}_{\bf 0}(t)$
is the reduced system density operator,
$\rho^{(n)}_{\bf j}(t)\equiv \rho^{(n)}_{j_1\cdots j_n}(t)$,
as specified in \Eq{DDO_def},
engages an \emph{ordered} set of $n$ \emph{irreducible} dissipatons.
Evidently, $\rho^{(n+1)}_{j{\bf j} }\equiv \rho^{(n+1)}_{jj_1\cdots j_n}
=(-)^n\rho^{(n+1)}_{{\bf j}j}$.
Denote also $\rho^{(n-1)}_{{\bf j}^-_r}
\equiv \rho^{(n+1)}_{j_1\cdots j_{r-1}j_{r+1}\cdots j_n}$.
The irreducible notation enables
the generalized Wick's theorem the
expression \cite{Yan14054105,Jin15234108,Yan16110306},
\begin{align}\label{Wick}
&\quad\, \text{tr}_{\B}\left[\big(\hat f_{j_n}\!\cdots\!\hat f_{j_1}\big)^{\circ}
    \hat f_j\rho_{\rm tot}(t)\right]   \nl
&=\rho^{(n+1)}_{j{\bf j}}
    + \sum_{r=1}^n (-)^{r-1} \La\hat f_{j_r}\hat f_j\Ra^{\greater}_{\B}
   \rho^{(n-1)}_{{\bf j}^{-}_r}.
\end{align}
Evidently, this can be used in evaluating
the effect of $H_{\SB}$--action on the specified
DDO, $\rho^{(n)}_{{\bf j}}(t)$.
Dissipaton algebra includes also
the generalized diffusion equation that
treats the effect of $H_{\B}$--action.

 Now, with the Liouville-von Neumann equation,
$\dot{\rho}_{\rm tot}(t)=-i[H_{\tS}+H_{\B}+H_{\SB},{\rho}
_{\rm tot}(t)]$, for the total density operator
in \Eq{DDO_def}, the aforementioned disspaton algebra
readily leads to \cite{Yan14054105,Jin15234108,Yan16110306}
\begin{align}\label{DEOM}
  \dot\rho^{(n)}_{\bf j}(t)&=-\bigg(i{\cal L}_{\tS}
  +\sum_{r=1}^n \gamma_{j_r}\bigg)\rho^{(n)}_{\bf j}(t)
  -i\sum_{j} {\cal A}_{\bar j}\rho^{(n+1)}_{{\bf j}j}(t)    \nl
&\quad
  -i \sum_{r=1}^n (-)^{n-r}{\cal C}_{j_r}\rho^{(n-1)}_{{\bf j}^-_r}(t).
\end{align}
While ${\cal L}_{\tS}\,(\cdot)=[H_{\tS},\,(\cdot)]$,
the Grassmannian superoperators, ${\cal A}_{\bar j}\equiv {\cal A}^{\bar\sigma}_{\alpha u\kappa} = {\cal A}^{\bar\sigma}_{u}$
and ${\cal C}_{j}\equiv {\cal C}^{\sigma}_{\alpha u\kappa}$,
are defined via 
\be\label{calAC}
\begin{split}
 {\cal A}^{\sigma}_{u} \Opm &\equiv
    a^{\sigma}_{ u}\Opm \pm \Opm \hat a^{\sigma}_{u}
 \equiv \big[\hat  a^{\sigma}_{u},\Opm\big]_\pm \, ,
\\
 {\cal C}^{\sigma}_{\alpha u\kappa} \Opm  &\equiv
  \sum_{v} \big(\eta^{\sigma}_{\alpha uv\kappa}\hat  a^{\sigma}_{v}\Opm
  \mp \eta^{\bar \sigma\,{\ast}}_{\alpha uv\kappa}\Opm \hat a^{\sigma}_{\kappa}\big).
\end{split}
\ee
Here, $\Opm$ is an arbitrary operator,
with even ($+$) or odd ($-$) fermionic parity,
such as $\rho^{(2m)}$ or $\rho^{(2m+1)}$, respectively.
Throughout this work, we adopt units of $e=\hbar=1$
for the electron charge and the Planck constant.

The DEOM theory, \Eqs{wtiF_f}--(\ref{DEOM}), describes both the
reduced system and hybrid bath dynamics.
The underlying DEOM--space quantum mechanics \cite{Yan16110306}
is a mathematical isomorphism of the conventional
Hilbert/Liouville--space
formulations.
It supports accurate evaluations of
the expectation values and correlation functions of
the type of $\hat A=\hat Q_{\tS}\hat F_{\B}$ operators,
including the cases of $\hat A=\hat Q_{\tS}$
and $\hat A=\hat F_{\B}$.
More specific, the system $\hat Q_{\tS}$
is arbitrary, such as the combination of
the creation (annihilation) operators $\hat a^{\sigma}_u$.
The bath one belongs to the hybridized set,
i.e., $\hat F_{\B}\in \{\hat F^{\pm}_{\alpha u}\}$.
Apparently, the type of $\hat A=\hat Q_{\tS}\hat F_{\B}$
includes the lead--specified transport current operator,
\be\label{hatI_alpha}
\hat I_{\alpha} = -\frac{\partial\hat N_{\alpha}}{\partial t}
 =-i\sum_u \big(\hat  a^{+}_u \hat F^-_{\alpha u}
   -\hat F^{+}_{\alpha u}\hat  a^-_u \big).
\ee
It is noticed that in general correlation functions
can be expressed in the form of
augmented expectation values; see \Eq{corr_hilbert} below.

 Let us start with time--dependent expectation values.
For a system dynamical operator it is
directly given by ${\rm Tr}[\hat Q_{\tS}\rho_{\rm tot}(t)]
={\rm tr}_{\tS}[\hat Q_{\tS}\rho_{\tS}(t)]$,
with $\rho_{\tS}(t)\equiv {\rm tr}_{\B} \rho_{\rm tot}(t)$
and ${\rm tr}_{\tS}$ being the trace over the system--subspace.
The average transient transport current,
$I_{\alpha}(t)\equiv {\rm Tr}[\hat I_{\alpha}\rho_{\rm tot}(t)]$,
for \Eq{hatI_alpha},
is evaluated by using the generalized Wick's theorem, \Eq{Wick}.
We obtain
\be\label{curr}
  I_{\alpha}(t)  = {\rm Tr}_{\T}\big[\hat I_{\alpha}\rho_{\rm tot}(t)\big]
= -i\! \sum_{j_{\alpha}\in j}
  {\rm tr}_{\tS}\!\big[\ti a_{\bar j}\rho^{(1)}_{j}(t)\big],
\ee
where $\ti a_{\bar j}\equiv \ti a^{\bar \sigma}_{\alpha u k}
=\bar \sigma\hat a^{\bar \sigma}_{u}$
and $j_{\alpha}\equiv \{ \sigma u k\}\in j\equiv\{\sigma\alpha u k\}$.

 On the other hand, the steady--state correlation functions
can generally be expressed in the form of expectation values as
\be\label{corr_hilbert}
\la \hat A(t) \hat B(0)\ra={\rm Tr}\big[\hat A\rho_{\rm tot}(t;\hat B )\big].
\ee
Here, $\hat O(t)\equiv e^{iH_{\rm tot}t}\hat O e^{-iH_{\rm tot}t}
=\hat O e^{-i{\cal L}_{\rm tot}t}$ the Heisenberg picture,
whereas
 $\rho_{\rm tot}(t;\hat B )\equiv
e^{-i{\cal L}_{\rm tot}t}(\hat B\rho_{\rm tot}^{\rm st})$
the Schr\"{o}dinger picture.
In the DEOM--space evaluation,
the steady--state total system--and--bath composite
$\rho^{\rm st}_{\rm tot}$ maps
to the steady--state DDOs, $\{\rho^{(n);{\rm st}}_{\bf j}\}$,
via \Eq{DDO_def}.
These are steady--state solutions to \Eq{DEOM}
which can readily be evaluated via, for instance,
the self-consistent iteration approach \cite{Zha17044105}.
Next, $\rho_{\rm tot}(t=0;\hat B )
 =\hat B\rho_{\rm tot}^{\rm st}$ maps to
$\{\rho^{(n)}_{\bf j}(t=0;\hat B )\}$, which
can be identified by using \Eqs{DDO_def} and (\ref{Wick}).
We then evaluate
$\rho_{\rm tot}(t;\hat B)\rightarrow \{\rho^{(n)}_{\bf j}(t;\hat B )\}$
via \Eq{DEOM}, and the correlation function via \Eq{corr_hilbert}.
The above mapping algorithm does exist whenever $\hat A$
and $\hat B$ belong to the aforementioned
$(\hat Q_{\tS}\hat F_{\B})$--type of the dynamical operators.
These include the DOS of the impurity,
\be
  A_u(\omega)=\frac{1}{2\pi}\!\int^\infty_{-\infty} \!\d t\,
 e^{i\omega t}\la \{\hat a_u(t),\hat a^\dg_u(0)\}\ra,
\ee
and the nonsymmetrized current noise spectrum,
\be\label{Sw_It}
  S_{\alpha\alpha'}(\omega)= \int_{-\infty}^{\infty}\!\!\d t\,
  e^{i\omega t} \La \delta{\hat I}_\alpha(t)\delta{\hat I}_{\alpha'}(0)\Ra.
\ee
Here, $\delta{\hat I}_\alpha\equiv{\hat I}_\alpha-I^{\rm st}_{\alpha}$,
with $I^{\rm st}_{\alpha}\equiv \la\hat I_{\alpha}\ra$
being the stationary current.
In contrast to the symmetrized one,
$S^{{\rm sym}}_{\alpha\alpha'}(\omega)
=S_{\alpha\alpha'}(\omega)
+S_{\alpha'\alpha}(-\omega)$,
the asymmetric $S_{\alpha\alpha'}(\omega)$
is directly related to experiments,
with $\omega>0$ and $<0$
corresponding to energy
absorption and emission processes,
respectively \cite{Eng04136602,Rot09075307,Yan14115411,
Bas10166801,Bas12046802,Del18041412,Moc11201303,%
Mul13245115,Cre18107702,Jin15234108}.
Further details of DEOM--space quantum mechanics can be found in
Ref.\,\onlinecite{Yan16110306}.

 Apparently, the DEOM--space evaluations
on expectation values and correlation functions
cover also that of the net circuit current,
\be\label{It}
 \hat I = a\hat I_{\rm L}-b\hat I_{\rm R}.
\ee
Its noise spectrum is
  $S(\omega)= \int_{-\infty}^{\infty} \!dt\,
  e^{i\omega t} \La \delta{\hat I}(t)\delta{\hat I}(0)\Ra$,
with $\delta{\hat I}\equiv{\hat I}-I^{\rm st}$.
The junction capacitance parameters are
$a=\Gamma_{\rm R}/\Gamma$ and
$b=\Gamma_{\rm L}/\Gamma$,
with $\Gamma=\Gamma_{\rm L}+\Gamma_{\rm R}$
being the total reservoirs coupling
strength \cite{Bla001,Eng04136602,Del18041412}.

 It is worth noting that the DEOM theory is a quasiparticle extension
to the well--established hierarchical equations of motion
formalism \cite{Jin08234703}.
 The latter consists only of \Eq{DEOM}
that has been demonstrated an efficient and universal
method for strongly correlated
quantum impurity systems
\cite{Zhe121129,Li12266403,Zhe13086601,Hou15104112,Ye16608}.
The reduced system density operator is just
$\rho_{\tS}(t) \equiv {\rm tr}_{\B}\rho_{\rm tot}(t)= \rho^{(0)}(t)$.
 All $\big\{\rho^{(n\geq1)}_{\bf j}\big\}$
are also physically well--defined DDOs, \Eq{DDO_def},
for entangled system--bath dynamics.
DEOM is naturally a nonperturbative many-particle theory and
is formally exact when
$n_{\rm max}=2N_{ \sigma}N_u$ \cite{Han18234108},
with $N_{u}$ being the number of spin--orbital states,
and $N_{ \sigma}=2$ being the two signs of $\sigma=+$ and $-$.
As an efficient and universal numerical method \cite{Jin08234703,Zhe121129,
Zhe13086601,Li12266403,Hou15104112,Ye16608},
DEOM converges rapidly and uniformly with increasing
the truncated tier level, $L=n_{\rm trun}$,
by setting all $\rho^{(n>L)}_{\bf j}=0$,
at a sufficiently large $L$ which is often much less than the maximum tier,
$n_{\rm max}$.
%
The minimal truncation tier $L$ required to achieve convergence is closely dependent on the
configurations of system as well as bath and especially the temperature of the bath.
In practice, the convergence with respect to $L$
is tested case by case.
For the parameters exemplified in the present study of the
Kondo problems, the accurate evaluations of the truncation tier
probably need $L>5$ and it is time consuming.
The numerical calculations here is thus up to $L=4$ tier level.
The convergence calculations for $L>4$ would
 correct the quantities of the transient current and its noise spectrum, but
would not affect their main characteristics that we will discuss
in the present work.

\section{Current noise spectrum and transient current}
\label{thnum}

\subsection{Equilibirum Kondo regime}

  For illustrations below,
we set the parameters in the Kondo regime (unit of meV):
$\varepsilon=-0.6$, $U=1.6$, $\Gamma=0.2$
and $k_{\rm B} T=0.005$,
for the impurity dot energy level, Coulomb interaction,
coupling reservoirs strength and temperature, respectively.
Adopt a wide bandwidth with $W=50\,\Gamma$ for both electrodes.
All parameters are within the current experiments \cite{Hol01256802,Del18041412}.
Set also $\Gamma_{\rm L}=\Gamma_{\rm R}=\Gamma/2$,
so that $a=b=1/2$ for \Eq{It}.
We focus on the symmetrical bias voltage
$\mu_{\rm L}=-\mu_{\rm R}=V/2$ throughout the paper,
unless otherwise stated.


 It is worth noting that for the SAIM QD in study,
both  the single--occupation ($\varepsilon$)
and the double--occupation ($\varepsilon+U$) transport channels
are relevant. We focus on the
Kondo tunneling regime where the former lies below the Fermi energy
($\varepsilon<0$) and the latter
is above ($\varepsilon+U>0$).
These two are called the Hubbard resonances.
Moreover, we set a low temperature
for the formation of Kondo's singlet state(s),
located at the Fermi level(s),
in either the equilibrium ($V\equiv\mu_{\rm L}-\mu_{\rm R}=0$)
or the nonequilibrium ($V\neq 0$) scenario.

 We focus on the total circuit noise
spectrum, $S(\omega)$,
that would be readily accessible in experiments \cite{Bas10166801,Bas12046802,Del18041412}.
As the Kondo characteristics are concerned,
we have verified that individual
noise spectrum, $S_{\rm L\rm L}(\omega)$, $S_{\rm R\rm R}(\omega)$
and ${\rm Re}[S_{\rm L\rm R}(\omega)]$
(not shown below) is similar to $S(\omega)$ for symmetrical coupling \cite{Jin15234108,Cre18107702}.
To elaborate the underlying picture
we present also other closely related properties.
These include $\d S(\omega)/\d\omega$,
the dot DOS $A(\omega)=A_\up(\omega)=A_\down(\omega)$,
and the transient current spectrum ${\cal J}(\w)$ that will be specified later.

\begin{figure}
\includegraphics[width=1.0\columnwidth]{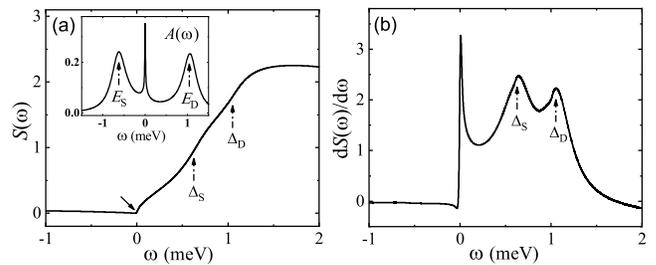} 
\caption{
The Kondo characteristics
on (a) circuit noise spectrum $S(\omega)$ (in $10^{-8}$ eA) and
(b) ${\rm d}S(\omega)/{\rm d}\omega$ (in $10^{-5}$A/V) at equilibrium
($V=0$ with $\mu^{\rm eq}_\alpha=0$).
The inset in (a) is the dot DOS
 $A(\omega)$.
}
\label{fig1}
\end{figure}

 Consider first the equilibrium ($V=0$) case,
i.e., \Fig{fig1},
with the evaluated $S(\w)$ and $\d S(\w)/\d\w$
being depicted in the panels (a) and (b),
respectively.
The DOS $A(\w)$ is given in the inset
of (a) for comparison.
The observations and elaborations are as follows.

Let us start with the Kondo characteristics here.
Evidently, in
$S(\w)$, $\d S(\w)/\d\w$ and $A(\w)$,
the equilibrium Kondo characteristics all appear at $\w=0$.
In particular, in the noise spectrum $S(\w)$
it exhibits an \emph{inflection} point (at $\w=0$),
which turns out to be a remarkable
\emph{Fano--type resonant peak}
in $\d S(\omega)/\d\omega$.
In contrast, the equilibrium Kondo resonance peak
in the DOS, $A(\w)$, is rather symmetric.
It is well--known that a perfectly symmetric $A(\w)$
goes with particle--hole symmetry \cite{Mei932601}.
On the other hand, the resulted
$S(\w)$ and $\d S(\w)/\d\w$
remain asymmetric, similar as above.

\subsection{Hubbard resonances and absorption mechanism
versus anti-Stokes co-tunneling resonance}

Turn to the two Hubbard resonances,
the single--occupation 
and the double--occupation transport resonances 
with energies of $\varepsilon<0$ and $\varepsilon+U>0$,
respectively.
While they are rather directly reflected in
the DOS $A(\w)$, these two states are manifested
in the noise spectrum, $S(\w)$,
via the \emph{absorption mechanism},
with the characteristics frequencies at
\be\label{nmks}
\Delta^{\alpha}_\text{\tiny{S}}
= \mu_{\alpha}-E_\text{\tiny{S}},~~~
\Delta^{\alpha}_\text{\tiny{D}}=
 E_\text{\tiny{D}}-\mu_{\alpha}.
\ee
Here,
$E_\text{\tiny{S}}$ and $E_\text{\tiny{D}}$
are the two Hubbard resonance energies of $A(\omega)$, as shown
by the two arrows in the inset of Fig.\,\ref{fig1}(a).
Note that $\mu^{\rm eq}_{\alpha}=0$ at equilibirum,
resulting in $\Delta^{\text{\tiny{L}}}_\text{\tiny{S}}
=\Delta^{\text{\tiny{R}}}_\text{\tiny{S}}=\Delta_\text{\tiny{S}}$
and
$\Delta^{\text{\tiny{L}}}_\text{\tiny{D}}
=\Delta^{\text{\tiny{R}}}_\text{\tiny{D}}=\Delta_\text{\tiny{D}}$,
 the dash arrows in Fig.\,\ref{fig1}.
Also note that
$E_\text{\tiny{S}}$ and
$E_\text{\tiny{D}}$
are not identical to the isolated energies in the dot,
but only approximately, i.e.,
$E_\text{\tiny{S}}\approx\varepsilon$ and
$E_\text{\tiny{D}}\approx\varepsilon+U$,
due to the renormalization \cite{Hau08}.

\begin{figure}
\includegraphics[width=1.04\columnwidth]{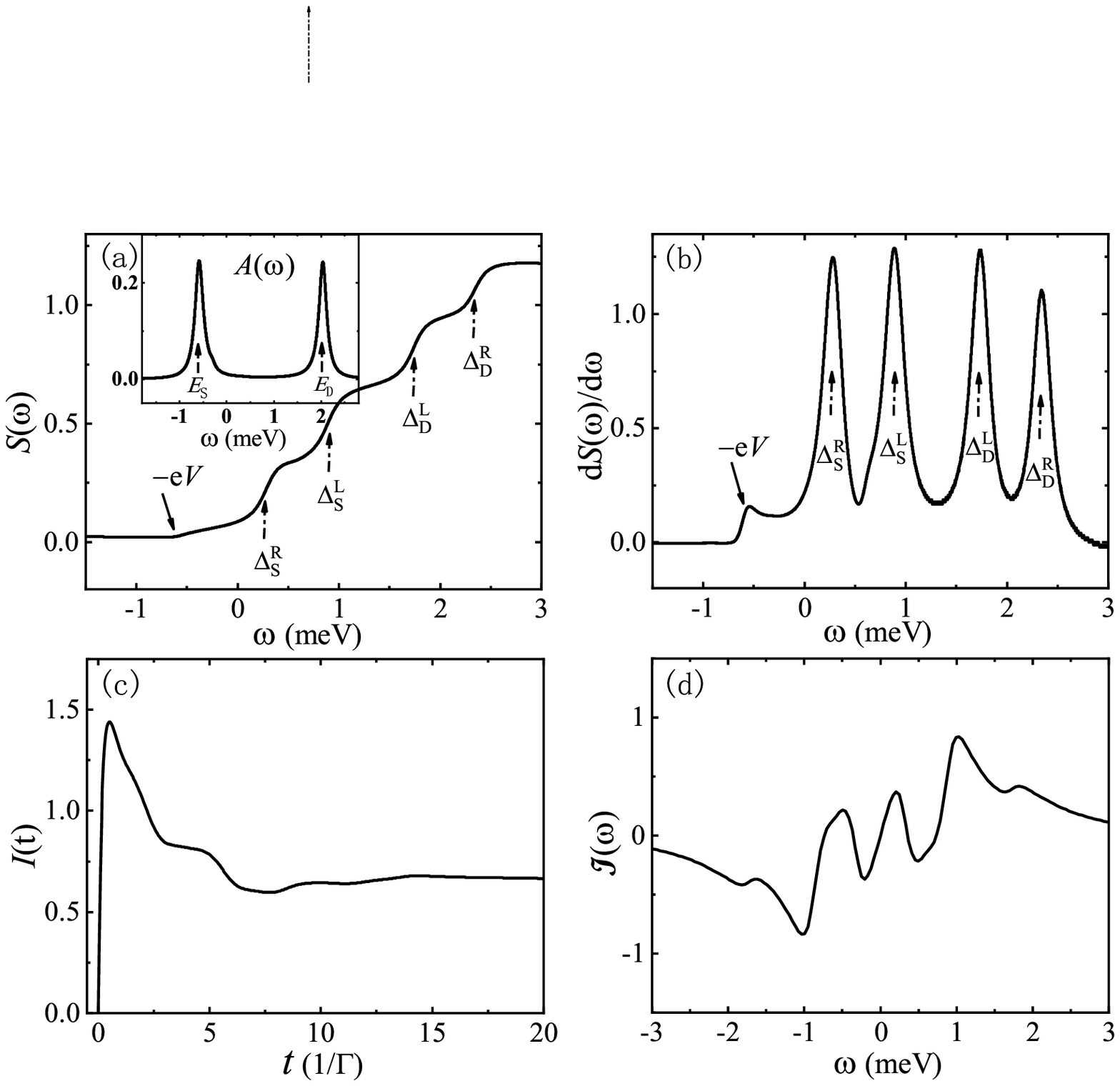} 
\caption{
The non-Kondo characteristics
on (a) circuit noise spectrum $S(\omega)$ (in $10^{-8}$ eA);
(b) ${\rm d}S(\omega)/{\rm d}\omega$ (in $10^{-5}$A/V);
(c) The real-time dynamics of the transport current $I(t)$ (in pA),
 with $I(t=0)=0$ being the equilibrium value,
before the bias voltage turns on;
(d) The sine transform of transient current,
 ${\cal J}(\omega)=\int^\infty_0\!{\rm d}t\,\sin(\omega t)\delta I(t)$,
 with $\delta I(t)=I(t)-I^{\rm st}$.
  The inset in (a) is the DOS
 $A(\omega)$.
The nonequilibrium
bias voltage is $V=0.6$
with $\mu_{\rm L}=-\mu_{\rm R}=V/2$. The
other parameters are: $\varepsilon=-0.6$, $U=2.6$,
$\Gamma_{\rm L}=\Gamma_{\rm R}=\Gamma/2=0.05$, and $k_{\rm B}T=0.02$.
}
\label{fig2}
\end{figure}

 To highlight the absorption mechanism,
we demonstrate the nonequilibrium
($V\neq 0$) noise spectrum in the non-Kondo regime
at an increased the temperature.
To have the absorptive feature more visible,
we also reduce the coupling strength ($\Gamma$)
and enhance the Coulomb interaction ($U$);
see the caption of Fig.\,\ref{fig2} for
the parameters.
The equilibrium characteristic at each
$\w=\Delta_\text{\tiny{S/D}}$
in \Fig{fig1} now splits into $\Delta^{\rm L}_\text{\tiny{S/D}}$
and $\Delta^{\rm R}_\text{\tiny{S/D}}$ in \Fig{fig2}.
The underlying mechanism is rather evident,
as the energy absorption involves
two sequential transport channels.
One goes by the tunneling of the electron in
the single--occupied state, with energy $E_\text{\tiny{S}}$,
to the $\alpha$-lead by absorbing the energy
$\Delta^{\alpha}_\text{\tiny{S}}$.
Another channel engages the electron in the $\alpha$-lead,
passing through the
double--occupation channel of $E_\text{\tiny{D}}$
by absorbing the energy $\Delta^{\alpha}_\text{\tiny{D}}$.
The opposite sequential processes
accompanied by energy emission do not happen.
Consequently, the absorption noise spectrum
$S(\omega)$ displays
rising steps around $\omega=\Delta^{\alpha}_\text{\tiny{S}}$
and $\Delta^{\alpha}_\text{\tiny{D}}$,
see \Fig{fig1}(a) and \Fig{fig2}(a).
These rising steps
in $S(\omega)$ are the sequential non-Markovian quasi steps
\cite{Jin11053704,Eng04136602,Rot09075307,Jin15234108}.
They are turn into Lorentzian-like peaks
 in $\d S(\omega)/\d\omega$ as plotted in Fig.\,\ref{fig1}(b)
and Fig.\,\ref{fig2}(b).

 Observed is also the \emph{anti-Stokes cotunneling resonance}
at the frequency, $\w=\Delta^{\text{\tiny{L}}}_\text{\tiny{D}}
-\Delta^{\text{\tiny{R}}}_\text{\tiny{D}}=-eV$.
As inferred from \Eq{nmks},
the double--occupation ($E_\text{\tiny{D}}$) transport channel
serves as the intermediate for
the coherent two--electron processes here.
This mechanism had been thoroughly analysed
in our pervious work \cite{Jin15234108},
with the lead--specific current noise spectrum
$S_{\alpha\alpha'}(\omega)$.
The observed anti-Stokes characteristic is particularly dominant
in  $S_{\rm L\rm R}(\omega)$. 
This highlights the underlying L-to-R (source-to-drain) cotunneling in nature.
The single--occupation ($E_\text{\tiny{S}}$)
below the Fermi surfaces would not contribute to
the coherent two--electron processes.
Moreover, the directionality of bias voltage suppresses
the inverse R-to-L Stokes cotunneling of $\Delta^{\text{\tiny{R}}}_\text{\tiny{D}}
-\Delta^{\text{\tiny{L}}}_\text{\tiny{D}}=+eV$.

It is well--known that
the DOS $A(\omega)$ does not involve
any cotunneling resonance.
On the other hand, nonequilibrium Kondo resonance
emerges distinguished peaks at Fermi energies in $A(\omega)$.
It is also noticed that
the nonequilibrium Kondo feature
in the current noise spectrum $S(\omega)$
appears at $\omega=\pm eV$
\cite{Moc11201303,Mul13245115,Cre18107702,Jin15234108}.

The main objective of this paper is to
elucidate the nonequilibirum Kondo mechanism
in the current noise spectrum $S(\omega)$.
For later comparison, we report the non-Kondo
 transient current
 $I(t)$ and its sine transform
 ${\cal J}(\omega)$,
   in Figs.\,\ref{fig2}(c) and (d), respectively.
   Their Kondo counterparts, Figs.\,\ref{fig3}(c) and (d),
   are remarkably different.

\subsection{Nonequilibrium Kondo regime}

\begin{figure}
\includegraphics[width=1.0\columnwidth]{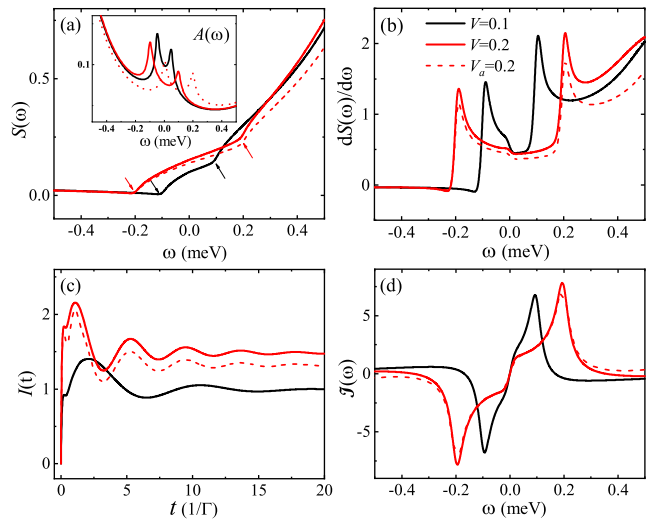} 
\caption{(Color online)
The numerical results of the Kondo characteristics
on (a) circuit noise spectrum $S(\omega)$ (in $10^{-8}$ eA);
(b) ${\rm d}S(\omega)/{\rm d}\omega$ (in $10^{-5}$A/V);
(c)The real-time dynamics of the transport current $I(t)$ (in pA);
(d) The sine transform of transient current,
 ${\cal J}(\omega)=\int^\infty_0\!{\rm d}t\,\sin(\omega t)\delta I(t)$.
  The inset in (a) is the DOS
 $A(\omega)$.
We adopt the
values of bias voltage (in meV):
$V=0.1$ (black) and $0.2$ (red).
The red-dashed curve is for the asymmetrical bias voltage
denoted by $V_a=0.2$ with $\mu_{\rm L}=0.2$meV
and $\mu_{\rm R}=0$.
}
\label{fig3}
\end{figure}

 Figure \ref{fig3}(a) reports the evaluated
 current noise spectrum $S(\omega)$,
at different values of the applied bias voltage.
Comparing to its equilibrium counterpart, \Fig{fig1}(a),
 the applied bias voltage
splits the Kondo characteristic, from
the single inflection point 
at $\omega=0$,
into two asymmetric upturns around
$\omega=\pm|\mu_{\rm L}-\mu_{\rm R}|=\pm eV$.
These differ also from the nonequilibrium
Kondo characteristic in the DOS $A(\w)$, the inset of \Fig{fig3}(a),
with the peaks at individual $\mu_{\rm L}$ and $\mu_{\rm R}$
 \cite{Mei932601,Leb01035308}.
The two asymmetric upturns in $S(\w)$
turn into two remarkable peaks in
$\d S(\omega)/\d\omega$,
at $\omega=\pm eV$, as plotted in \Fig{fig3}(b).
%

To highlight the fact that the 
Kondo characteristic in
 $S(\omega)$ is concerned only with
the difference between two Fermi energies,
we consider also the case
of the asymmetrical bias voltage.
The red-dash curves in Fig.\ref{fig3}
report the case of $\mu_{\rm L}=0.2$\,meV and $\mu_{\rm R}=0$.
The Kondo chareristics appears at the same frequencies as
    the symmetrical bias voltage case
  (red-solid) with the same Fermi energies difference.

The scenario in $S(\omega)$ differs from
 that in $A(\omega)$. The latter is
 depicted in the inset of \Fig{fig3}(a),
where the Kondo resonance follows the individual $\mu_{\rm L}$ and $\mu_{\rm R}$.
This is related to the formation of Kondo
singlet at the Fermi surfaces \cite{Mei932601,Leb01035308}.
While the DOS $A(\omega)$ reflects the structure information,
the current noise spectrum is
related to not only the structure but also the transport current
dynamics.
The transient current $I(t)$ reported in \Fig{fig3}(c)
displays Kondo oscillation dynamics as consistent with
the previous work \cite{Che15033009}.
This Kondo oscillation comes from
the Rabi interference between two Kondo
resonance transport channels of $\mu_{\rm L}$
and $\mu_{\rm R}$.
This type of interference does not exist in the non-Kondo regime
as exemplified in  \Fig{fig2}(c).
Further depicting the sine transform of the transient current
 ${\cal J}(\omega)$ in Fig.\ref{fig3}(d),
it exhibits the dip and peak at $\omega=-eV$ and $\omega=eV$, respectively.
This feature is also remarkably different from that of the non-Kondo regime
as shown in Fig.\ref{fig2}(d).
The Kondo oscillation frequency of
the transient current
is $|eV|=|\mu_{\rm L}-\mu_{\rm R}|$
and independent of the specific Fermi energies
(see the red-solid versus the red-dash).
Note that the appearance of the dip/peak at $\omega=\pm eV$
 comes from the nature of the sine transformation, i.e.,
${\cal J}(\omega)=-{\cal J}(-\omega)$.
We now conclude that
the Kondo characteristic in the noise spectrum
reflects the Rabi interference of the transport current dynamics.
The emergence of Kondo feature located
at $\omega=\pm eV$ in $S(\omega)$ contains the information of the
Kondo oscillation frequency $|eV|$.
Evidently, there is a bridge
between the nonequilibrium Kondo noise
and the transient current.
 By comparing to the non-Kondo regime,
\Fig{fig2}(b) and (d),
emerged in the Kondo regime, \Fig{fig3}(b) and (d) are
also minor but distinguished
inflections near zero-frequency.
Comparing further to the equilibrium Kondo counterpart, \Fig{fig1},
we could conclude that the observed
inflection characteristic in $\w\in [-eV, +eV]$
is a sort of \emph{Kondo--Fano interference}.
This engages both Kondo characteristics
at $-eV$ and $+eV$, which differs from
the anti-Stokes cotunneling feature at $-eV$ only,
as seen in \Fig{fig2}(a) and (b).

 As the Kondo feature is concerned,
$S(\w)$ and $\d S(\w)/\d\w$ in the emission ($\w<0$) region is
more distinguishable than the absorption ($\w>0$)
region; see Fig.\,\ref{fig3}.
The spectroscopic information in $\w>0$
is complicated by the sequential tunneling resonances
at $\w=\Delta^{\alpha}_{\tS}$ and $\Delta^{\alpha}_\text{\tiny D}$
of \Eq{nmks}.
The observation here goes often in
practical reporting experimental
results of the Kondo noise spectrum
in the $\w<0$ region \cite{Bas12046802,Del18041412}.
The visualization in the $\w>0$ region
would be possible when there is
clear separation between the Kondo
and the sequential resonances \cite{Bas10166801}.

\section{Summary}
\label{thsum}

In summary, we have investigated the circuit current
noise spectrum through an Anderson impurity quantum dot
and underlying transient dynamics
in the Kondo regime.
Based on the DEOM evaluations, we
first demonstrate the equilibrium case,
where the Kondo resonance peak
in the DOS, $A(\w)$, is rather symmetric
around $\w=0$.
The responding Kondo characteristic in
the noise spectrum $S(\w)$
exhibits an \emph{inflection} point,
and that in $\d S(\omega)/\d\omega$,
turns out to be a remarkable
\emph{Fano--type resonant peak},
at $\w=0$.

 It is well--known that the noise spectrum can be tuned
by bias voltage.
The related absorption mechanism,
with the electron sequential tunneling
resonances at $\Delta^{\alpha}_\text{\tiny{S,D}}$
of \Eq{nmks}.
On the other hand, $A(\w)$ directly
reflects the two Hubbard resonances at $E_\text{\tiny{S,D}}$
that are independent of the bias voltage.
The tunneling resonances
at $\Delta^{\alpha}_\text{\tiny{S,D}}$
are the non-Markovian quasi-steps
and Lorentzian-like peaks,
in $S(\w)$ \cite{Jin11053704,Eng04136602,Rot09075307} and $\d S(\omega)/\d\omega$,
respectively.

 We then study the nonequilibrium Kondo characteristics.
The applied bias voltage
splits the characteristic in $S(\w)$, from
the single inflection point 
at $\omega=0$,
into two asymmetric upturns around
$\omega=\pm|\mu_{\rm L}-\mu_{\rm R}|=\pm eV$.
Meanwhile, in $\d S(\omega)/\d\omega$
the Kondo features are two remarkable peaks.
We demonstrate that the observed absorptive/emissive Kondo resonance
is concerned only with $eV=\mu_{\rm L}-\mu_{\rm R}$.
This differs from the DOS $A(\omega)$,
where the Kondo resonance peaks
reflect rather the structure information
for the shifted Fermi energies of
$\mu_{\rm L}$ and $\mu_{\rm R}$.
In other words, the DOS describes the formation of Kondo resonance
singlet states on individual nonequilibrium Fermi surfaces.

 Current noise spectrum is
related to not only the structure
but also the transport current dynamics.
Further demonstrations include also
the transient circuit current.
Evidently, the nonequilibrium Kondo features
at $\omega=\pm eV$ originate from the
Kondo oscillation of the transport current.
Moreover, we compare the noise spectra
between the non-Kondo and Kondo regimes
and between the equilibrium and nonequilibrium cases.
The observed overall inflection characteristics within
$\w\in [-eV, eV]$
indicate a sort of \emph{Kondo--Fano interference}.
This engages both Kondo characteristics
at $-eV$ and $+eV$, which differs from
the anti-Stokes cotunneling feature at $-eV$ only.
The emission noise Kondo feature is often
more distinguishable than the absorption, as the
latter would be contaminated by the sequential tunneling signals.
This work is closely related to
the experiments \cite{Bas10166801,Bas12046802,Del18041412}.
It could be anticipated that the
present results can be readily demonstrated in the current experiments.

\acknowledgments
Support from the Natural Science Foundation
of China (Nos.\ 11675048, 21633006 \& 11447006)
is gratefully acknowledged.

\end{document}